# Universal Voting Protocol Tweaks to Make Manipulation Hard[*]

**Vincent Conitzer** and **Tuomas Sandholm**
Carnegie Mellon University
Computer Science Department
5000 Forbes Avenue
Pittsburgh, PA 15213, USA
{conitzer,sandholm}@cs.cmu.edu

## Abstract

Voting is a general method for preference aggregation in multiagent settings, but seminal results have shown that all (nondictatorial) voting protocols are manipulable. One could try to avoid manipulation by using voting protocols where determining a beneficial manipulation is hard computationally. A number of recent papers study the complexity of manipulating existing protocols. This paper is the first work to take the next step of *designing new protocols that are especially hard to manipulate*. Rather than designing these new protocols from scratch, we instead show how to *tweak* existing protocols to make manipulation hard, while leaving much of the original nature of the protocol intact. The tweak studied consists of adding one elimination preround to the election. Surprisingly, this extremely simple and universal tweak makes typical protocols hard to manipulate! The protocols become NP-hard, #P-hard, or PSPACE-hard to manipulate, depending on whether the schedule of the preround is determined before the votes are collected, after the votes are collected, or the scheduling and the vote collecting are interleaved, respectively. We prove general sufficient conditions on the protocols for this tweak to introduce the hardness, and show that the most common voting protocols satisfy those conditions. These are the first results in voting settings where manipulation is in a higher complexity class than NP (presuming PSPACE $\neq$ NP).

## 1 Introduction

Often, a group of agents has to make a common decision, yet they have different preferences about which decision is made. Thus, it is of central importance to be able to aggregate the preferences, that is, to make a socially desirable decision as to which *candidate* is chosen from a set of candidates. Such candidates could be potential presidents, joint plans, allocations of goods or resources, etc. Voting is the most general preference aggregation scheme, and has been used in several multiagent decision making problems in AI, such as collaborative filtering (e.g. [Pennock *et al.*, 2000]) and planning among automated agents (e.g. [Ephrati and Rosenschein, 1991; Ephrati and Rosenschein, 1993]).

A key problem voting mechanisms are confronted with is that of *manipulation* by the voters. An agent is said to vote strategically when it does not rank the alternatives according to its true preferences, but differently so as to manipulate the outcome to be more favorable to the agent. For example, if an agent prefers Nader to Gore to Bush, but knows that Nader has too few other supporters to win, while Gore and Bush are close to each other, the agent would be better off by declaring Gore as its top candidate. Manipulation is an undesirable phenomenon. For one, because social choice schemes are tailored to aggregate preferences in a socially desirable way, and if the agents reveal their preferences insincerely, a socially undesirable candidate may be chosen.

A seminal negative result, the *Gibbard-Satterthwaite theorem*, shows that if there are three or more candidates, then in any nondictatorial voting scheme, there are preferences under which an agent is better off voting strategically [Gibbard, 1973; Satterthwaite, 1975]. (A voting scheme is called dictatorial if one of the voters dictates the social choice no matter how the others vote). In automated group decision making where the voters are software agents, the manipulability of protocols is even more problematic, for at least two reasons. First, the algorithms they use to decide how to vote must be coded explicitly. Given that the voting algorithm needs to be designed only once (by an expert), and can be copied to large numbers of agents (even ones representing unsophisticated human voters), it is likely that rational strategic voting will increasingly become an issue, unmuddied by irrationality, emotions, etc. Second, software agents have more computational power and are more likely to find effective manipulations.

We take the following tack toward avoiding manipulation: *ensuring that finding a beneficial manipulation is so hard computationally that it is unlikely that voters will be able to manipulate*. So, unlike in most of computer science, here high computational complexity is a desirable property. The harder it is to manipulate, the better.

[*]The material in this paper is based upon work supported by the National Science Foundation under CAREER Award IRI-9703122, Grant IIS-9800994, ITR IIS-0081246, and ITR IIS-0121678.

Prior work on the complexity of manipulating elections has focused on existing protocols [Bartholdi *et al.*, 1989; Bartholdi and Orlin, 1991; Conitzer and Sandholm, 2002a]. This paper is the first to take the next step of designing new protocols that are especially hard to manipulate. Rather than designing these protocols from scratch, we show how to tweak existing protocols to make manipulation computationally much more difficult, while leaving much of the original nature of the protocol intact, for the following reasons:

- Results on the computational complexity induced by a tweak typically apply to a large family of protocols.
- Many of the original protocol's nice theoretical properties are preserved by the tweak.
- In practice, it will be much easier to replace a currently used protocol with a tweaked version of it, than with an altogether new protocol.

The type of tweak we study in this paper is the following. All the candidates are paired in a *preround*; of each pair of candidates, only the winner of their *pairwise election* survives. (The winner of the pairwise election between two candidates is the candidate that is ranked above the other more often in the votes.) After the preround, the original protocol is executed on the remaining candidates. The *schedule* of the preround (i.e., who faces who) can be determined before the votes are collected; after the votes are collected; or while the votes are collected (the processes are interleaved). We study these three cases in Sections 4, 5, and 6, respectively. [1]

## 2 Definitions

### 2.1 Elections and voting protocols

An *election* consists of a set of candidates $C$; a set of voters $V$; and a protocol for deciding on a winner $w \in C$ given all the voters' votes. (Here, a vote is a total ordering of the candidates.) A *deterministic* protocol is a function from the set of all combinations of votes to $C$. (All our results hold even for unweighted voters, so in this paper this function will always treat the voters symmetrically.) A *randomized* protocol is a function from the set of all combinations of votes to probability distributions over $C$. An *interleaved* protocol is a procedure for alternating between collecting parts of the voters' votes (e.g. whether they prefer candidate $a$ to candidate $b$) and drawing and publishing random variables (such as parts of the schedule for an election), together with a function from the set of all combinations of votes and random variables to $C$. (Collecting only parts of the voters' preferences is also known as *elicitation*.)

The standard definitions of most voting protocols allow for the possibility of ties between candidates, in which case a tie-breaking rule is required to fully specify the protocol. All our results hold for any tie-breaking rule, so we do not need to specify such rules here.

In this paper we apply our technique to the most common voting protocols (in these definitions, whenever points are defined, the candidate with the most points wins):

- *Plurality.* A candidate receives 1 point for every voter that ranks it first. (Thus, the voters effectively only vote for one candidate.)
- *Borda.* For each voter, a candidate receives $m-1$ points if it is the voter's top choice, $m-2$ if it is the second choice, ..., 0 if it is the last.
- *Maximin.* A candidate's score in a *pairwise election* is the number of voters that prefer it over the opponent. A candidate's number of points is the lowest score it gets in any pairwise election.
- *Single Transferable Vote (STV).* The winner determination process proceeds in rounds. In each round, a candidate's score is the number of voters that rank it highest among the remaining candidates, and the candidate with the lowest score drops out. The last remaining candidate wins. (A vote "transfers" from its top remaining candidate to the next highest remaining candidate when the former drops out.)

### 2.2 Preround

The tweaks we study in this paper all involve the addition of a preround. We will now define how this works.

**Definition 1** *Given a protocol P, the new protocol obtained by adding a preround to it proceeds as follows:*

1. *The candidates are paired. If there is an odd number of candidates, one candidate gets a bye.*
2. *In each pairing of two candidates, the candidate losing the pairwise election between the two is eliminated. A candidate with a bye is never eliminated.*
3. *On the remaining candidates, P is executed to produce a winner. For this, the implicit votes over the remaining candidates are used. (For example, if a voter voted $a \succ b \succ c \succ d \succ e$, and $b$ and $c$ were eliminated, the voter's implicit vote is $a \succ d \succ e$.)*

*The pairing of the candidates is also known as the* schedule *for the preround. If the schedule is decided and published before the votes are collected, we have a* deterministic preround *($DPRE$), and the resulting protocol is called $DPRE+P$. If the schedule is drawn completely randomly after the votes are collected, we have a* randomized preround *($RPRE$), and the resulting protocol is called $RPRE + P$. Finally, if the votes are elicited incrementally, and this elicitation process is interleaved with the scheduling-and-publishing process (which is again done randomly), as described in detail in Section 6, we have an* interleaved preround *($IPRE$), and the resulting protocol is called $IPRE + P$.*

---

[1] The high complexity results obtained in this paper are dependent on the number of candidates growing. This places them in line with all the early results in this area [Bartholdi *et al.*, 1989; Bartholdi and Orlin, 1991], but in contrast with more recent results [Conitzer and Sandholm, 2002a] that show high complexity of manipulation with a constant number of candidates for some protocols. Having high complexity of manipulation occur with a constant number of candidates already is certainly preferable to having it occur only with a growing number. On the other hand, unlike in that paper, the results here hold even when the voters all have equal weight, and even when manipulation is attempted by an individual rather than a coalition, making the results in this paper stronger in that sense, so there is a tradeoff.

## 2.3 Manipulation

We now define the computational problem of manipulation. Because all our hardness results hold even when the voters are unweighted, only a single voter is trying to manipulate, and all the other voters' votes are known to the manipulator, we will only define this simple setting here. Any hardness results in this simple setting immediately imply hardness in all more general settings.

**Definition 2 (CONSTRUCTIVE-MANIPULATION)**
*We are given a protocol $P$, a candidate set $C$, a preferred candidate $p$, and a set of votes $S$ corresponding to all the other voters' votes. The manipulator has yet to decide on its vote, and wants to make $p$ win. Then the constructive manipulation question is:*

- *(For deterministic protocols) Can the manipulator cast its vote to make $p$ win under $P$?*
- *(For randomized protocols) Can the manipulator cast its vote to make the probability of $p$ winning under $P$ at least some given $k \in [0, 1]$?*
- *(For interleaved protocols) Given the initial random choices (if any) by the protocol, is there a contingency plan (based on the random decisions the protocol takes between eliciting parts of the votes) for the manipulator to answer the queries to make the probability of $p$ winning under $P$ at least some given $k \in [0, 1]$?*

## 3 Complexity of manipulating untweaked protocols

In this section, we briefly review the complexity of manipulating voting protocols, as a benchmark for our results. CONSTRUCTIVE-MANIPULATION is in P for the Plurality, Borda, and Maximin voting protocols [Bartholdi *et al.*, 1989]. The only voting protocol for which CONSTRUCTIVE-MANIPULATION is known to be NP-hard is the STV protocol [Bartholdi and Orlin, 1991].[2]

## 4 NP-hardness when scheduling precedes voting

In this section, we examine the complexity induced by the preround when the voters know the schedule before they vote.

### 4.1 A sufficient condition for NP-hardness

We present a sufficient condition under which adding a preround with a preannounced schedule makes manipulation NP-hard. The condition can be thought of as an NP-hardness reduction template. If it is possible to reduce an arbitrary SAT instance to a set of votes satisfying certain properties under the given voting protocol, that protocol—with a preround—is NP-hard to manipulate.

**Theorem 1** *Given a voting protocol $P$, suppose that it is possible, for any Boolean formula $\phi$ in conjunctive normal form*

---
[2]CONSTRUCTIVE-MANIPULATION is NP-hard also for the *Second Order Copeland* protocol [Bartholdi *et al.*, 1989], but the hardness is driven solely by the tie-breaking rule.

*(i.e., a SAT instance), to construct in polynomial time a set of votes over a candidate set containing at least $\{p\} \cup C_L$ where $C_L = \{c_l : l \in L\}$ ($L$ is the set of literals $\{+v : v \in V\} \cup \{-v : v \in V\}$, where $V$ is the set of variables used in $\phi$), with the following properties:*

- *(Property 1a) If we remove, for each $v \in V$, one of $c_{+v}$ and $c_{-v}$, $p$ would win an election under protocol $P$ against the remaining candidates if and only if for every clause $k \in K$ (where $K$ is the set of clauses in $\phi$), there is some $l \in L$ such that $c_l$ has not been removed, and $l$ occurs in $k$. This should hold even if a single arbitrary vote is added.*
- *(Property 1b) For any $v \in V$, $c_{+v}$ and $c_{-v}$ are tied in their pairwise election after these votes.*

*Then CONSTRUCTIVE-MANIPULATION in $DPRE + P$ is NP-hard (and NP-complete if $P$ is deterministic and can be executed in polynomial time).*

**Proof**: Consider the following election under $DPRE + P$. Let the candidate set be the set of all candidates occurring in the votes constructed from $\phi$ (the "original candidates"), plus one dummy candidate for each of the original candidates besides those in $C_L$. To each of the constructed votes, add all the dummy candidates at the bottom; let the resulting set of votes be the set of the nonmanipulators' votes. A single manipulator's vote is yet to be added. Let the schedule for the preround be as follows: for each $v$, $c_{+v}$ and $c_{-v}$ face each other in the preround; and every other original candidate faces (and, because of the dummy candidates' position in the votes, defeats) a dummy candidate. Thus, the set of candidates that make it through the preround consists of, for each $v \in V$, one of $c_{+v}$ and $c_{-v}$; and all the other original candidates. The manipulator's vote will decide the winner of every $c_{+v}$ vs. $c_{-v}$ match-up, because by property 1b, all these pairwise elections are currently tied. Moreover, it is easy to see that the manipulator can decide the winner of each of these match-ups independently of how it decides the winners of the other match-ups. Thus, we can think of this as the manipulator giving the variables truth-values: $v$ is set to *true* if $c_{+v}$ survives, and to *false* if $c_{-v}$ survives. By property 1a it then follows that $p$ wins if and only if the manipulator's assignment satisfies all the clauses, i.e. is a solution to the SAT instance. Hence there is a successful constructive manipulation if and only if there is a solution to the SAT instance, and it follows that CONSTRUCTIVE-MANIPULATION in $DPRE + P$ is NP-hard. (It is also in NP if $P$ is deterministic and can be executed in polynomial time, because in this case, given a vote for the manipulator, it can be verified in polynomial time whether this vote makes $p$ win). ∎

### 4.2 Examples

We now show how to apply Theorem 1 to the well-known protocols we discussed, thus showing that each of these protocols—with a preround—is NP-hard to manipulate.

**Theorem 2** *There exists a reduction that satisfies properties 1a and 1b of Theorem 1 under the Plurality protocol.*

When it does not matter for our proofs whether a given vote is $a \succ b \succ c$ or $b \succ a \succ c$, we write $\{a, b\} \succ c$.

**Proof**: Given the formula $\phi$, let the candidate set be the minimally required candidates $\{p\} \cup C_L$, plus a set of candidates corresponding to the set of clauses $K$ of $\phi$, $C_K = \{c_k : k \in K\}$. Then, let the set of votes be as follows: $4|K| + 2$ votes ranking the candidates $p \succ C_L \succ C_K$; for each $k \in K$, $4|K|$ votes ranking the candidates $c_k \succ \{c_{cl} \in C_K : cl \neq k\} \succ C_L \succ p$; and for each $k \in K$, 4 votes ranking the candidates $\{c_l \in C_L : l \in k\} \succ c_k \succ \{c_l \in C_L : l \notin k\} \succ \{c_{cl} \in C_K : cl \neq k\} \succ p$. Additionally, we require that these votes are such that after counting them, for each $v \in V$, $c_{+v}$ and $c_{-v}$ are tied in their pairwise election, so that property 1b is satisfied. (This is possible because the total number of votes is even, and the majority of the votes do not yet have any restrictions on the order of the $C_L$.) We now show property 1a is satisfied. We first observe that regardless of which of the candidates corresponding to literals are removed, $p$ will get $4|K| + 2$ votes. Now, if for some $k \in K$, all the candidates $c_l$ with $l \in L, l \in k$ are removed, then $c_k$ will get at least $4|K| + 4$ votes and $p$ will not win. On the other hand, if for each $k \in K$, at least one candidate $c_l$ with $l \in k$ remains, then each of the $c_k$ will get precisely $4|K|$ votes. Because each remaining $c_l$ can get at most $4|K|$ votes as well, $p$ will win. In both cases there is a "margin" of at least 2, so a single additional vote will not change this. Thus, property 1a is satisfied. ∎

**Theorem 3** *There exists a reduction that satisfies properties 1a and 1b of Theorem 1 under the Borda protocol.*

**Proof**: Given the formula $\phi$, let the candidate set be the minimally required candidates $\{p\} \cup C_L$; plus a set of candidates corresponding to the set of clauses $K$ of $\phi$, $C_K = \{c_k : k \in K\}$, which we order in some arbitrary way to get $\{c_1, \ldots, c_{|K|}\}$. Let $M$ be the total number of candidates this defines. Then, let the set of votes be as follows: for every $c_i \in C_K$, $4M$ votes ranking the candidates $c_{i+1} \succ c_{i+2} \succ \ldots \succ c_{|K|} \succ p \succ c_1 \succ c_2 \succ \ldots \succ c_{i-1} \succ \{c_l \in C_L : l \in c_i\} \succ c_i \succ \{c_l \in L : l \notin c_i\}$; (here, the slight abuse of notation $l \in c_i$ means that $l$ occurs in the clause corresponding to $c_i$;) $4M$ votes ranking the candidates $c_1 \succ c_2 \succ \ldots \succ c_{|K|} \succ p \succ C_L$; one vote $c_1 \succ c_2 \succ \ldots \succ c_{|K|} \succ C_L \succ p$; one vote $c_{|K|} \succ c_{|K|-1} \succ \ldots \succ c_1 \succ C_L \succ p$; and finally, $4|K|M$ votes ranking the candidates $p \succ c_1 \succ c_2 \succ \ldots \succ c_n \succ C_L$, and $4|K|M$ votes ranking the candidates $c_n \succ c_{n-1} \succ \ldots c_1 \succ p \succ C_L$. Additionally, we require that these votes are such that after counting them, for each $v \in V$, $c_{+v}$ and $c_{-v}$ are tied in their pairwise election, so that property 1b is satisfied. (This is possible because the total number of votes is even, and the majority of the votes do not yet have any restrictions on the order of the $c_l$.) We now show property 1a is satisfied. It is easy to see that none of the $c_l$ can win, regardless of which of them are removed. Thus, we only need to consider the $c_i$ and $p$. The last $8|K|M$ votes will have no net effect on the relative scores of these candidates, so we need not consider these here. After the first $4(|K| + 1)M$ votes, any $c_k$ for which all the $c_l$ with $l \in k$ have been removed will be tied with $p$, and any other $c_k$ will be at least $4M$ points behind $p$. Finally, from the last remaining two votes, any $c_k$ ($k \in K$) will gain $2M - 2|V| - |K| - 1$ points on $p$. It follows that $p$ wins if and only if for every clause $k \in K$, there is some $l \in L$ with $l \in k$ such that $c_l$ has not been removed. In both cases there is a "margin" of at least $M - |V|$ points, so a single additional vote will not change this. Thus, property 1a is satisfied. ∎

**Theorem 4** *There exists a reduction that satisfies properties 1a and 1b of Theorem 1 under the Maximin protocol.*

**Proof**: Given the formula $\phi$, let the candidate set be the minimally required candidates $\{p\} \cup C_L$, plus a set of candidates corresponding to the set of clauses $K$ of $\phi$, $C_K = \{c_k : k \in K\}$. Then, let the set of votes be as follows: $8|K|$ votes ranking the candidates $p \succ C_L \succ C_K$, $8|K|$ votes ranking the candidates $C_L \succ C_K \succ p$, and $8|K|$ votes ranking the candidates $C_K \succ p \succ C_L$; $4|K|$ votes ranking the candidates $C_L \succ p \succ C_K$, $4|K|$ votes ranking the candidates $C_K \succ C_L \succ p$, and, for each $k \in K$, 4 votes ranking the candidates $p \succ \{c_{cl} \in C_K : cl \neq k\} \succ \{c_l \in C_L : l \in k\} \succ c_k \succ \{c_l \in C_L : l \notin k\}$; and finally, 2 votes ranking the candidates $p \succ C_K \succ C_L$, and 2 votes ranking the candidates $C_K \succ p \succ C_L$. Additionally, we require that these votes are such that after counting them, for each $v \in V$, $c_{+v}$ and $c_{-v}$ are tied in their pairwise election, so that property 1b is satisfied. (This is possible because the total number of votes is even, and the majority of the votes do not yet have any restrictions on the order of the $c_l$.) We now show property 1a is satisfied. Regardless of which of the candidates corresponding to literals are removed, $p$'s worst score in a pairwise election is against any of the $c_k$, namely $16|K| + 2$. Any $c_k$ for which all the $c_l$ with $l \in k$ have been removed will get its worst pairwise election score against any of the $C_L$, namely $16|K| + 4$. Finally, any other $c_k$ will get its worst pairwise election score against one of the $c_l$ with $l \in k$, namely, $16|K|$. It follows that $p$ wins if and only if for every clause $k \in K$, there is some $l \in k$ such that $c_l$ has not been removed. In both cases there is a "margin" of at least 2, so a single additional vote will not change this. Thus, property 1a is satisfied. ∎

**Theorem 5** *There exists a reduction that satisfies properties 1a and 1b of Theorem 1 under the STV protocol.*

**Proof**: We omit the proof due to limited space. ∎

**Theorem 6** *In any of $DPRE + Plurality$, $DPRE + Borda$, $DPRE + Maximin$, and $DPRE + STV$[3], CONSTRUCTIVE-MANIPULATION is NP-complete.*

**Proof**: NP-hardness is immediate from the previous theorems. The problem is in NP because these protocols can be

---

[3] The NP-completeness of manipulating $DPRE + STV$ is, in itself, not that interesting, because STV is already NP-hard to manipulate without the preround as we discussed. Nevertheless, our method highlights a different aspect of the NP-hardness of manipulating $DPRE + STV$. We build on this reduction later to prove PSPACE-hardness of manipulating STV with a preround when the scheduling of the preround is interleaved with the vote elicitation.

executed in polynomial time. ∎

In the next sections, we will raise the bar and bring the problem of manipulating elections to higher complexity classes by abandoning the assumption that the schedule for the preround should be known in advance.

# 5 #P-hardness when voting precedes scheduling

In this section, we will examine the complexity induced by the preround when the schedule is drawn completely (uniformly) randomly after all the votes have been collected.

## 5.1 A sufficient condition for #P-hardness

We present a sufficient condition for a voting protocol to become #P-hard[4] to manipulate in this setting. Again, this condition can be thought of as a reduction template. If it is possible to reduce an arbitrary PERMANENT instance to a set of votes satisfying certain properties under the given voting protocol, that protocol is #P-hard to manipulate when a randomized preround is added to it. (In the PERMANENT problem, we are given a bipartite graph $B$ with the same number of vertices $k$ in both parts, and are asked how many matchings there are. This problem is #P-complete [Valiant, 1979].)

**Theorem 7** *Given a voting protocol $P$, suppose that it is possible, for any bipartite graph $B$ with the same number of vertices $k$ in both parts (labeled $1$ to $k$ in one part, $k+1$ to $2k$ in the other), to construct in polynomial time a set of votes over the candidate set $\{c_1, \ldots, c_{2k}, p\}$ (where $c_i$ corresponds to vertex $i$ in $B$) with the following properties:*

- *(Property 2a) If we remove $k$ of the $c_i$, $p$ would win an election under protocol $P$ against the remaining $c_i$ if and only if the removed $c_i$ are exactly all the $c_i$ with $k+1 \leq i \leq 2k$;*
- *(Property 2b) $p$ loses its pairwise election against all $c_i$ with $k+1 \leq i \leq 2k$;*
- *(Property 2c) For any $1 \leq i \leq k$ and $k+1 \leq j \leq 2k$, $c_i$ defeats $c_j$ in their pairwise election if and only if in $B$, there is an edge between vertices $i$ and $j$.*
- *(Property 2d) All the previous properties still hold with any additional single vote.*

*Then CONSTRUCTIVE-MANIPULATION in $RPRE + P$ is #P-hard.*

**Proof**: Given the set of votes constructed on the basis of an arbitrary $B$, let us compute the probability that $p$ wins under the protocol $RPRE + P$ with only these votes. In the preround, there are $k$ matches and one bye. By property 2a, $p$ will win the election if and only if the $k$ candidates eliminated in this preround are precisely all the $c_i$ with $k+1 \leq i \leq 2k$. By property 2b, $p$ could not win a preround match against any of these, so $p$ will win the election if and only if it gets the bye, and each of the $c_j$ with $k+1 \leq j \leq 2k$ faces one of the $c_i$ with $1 \leq i \leq k$ that defeats it in the preround. Then, by property 2c, it follows that $p$ wins if and only if the preround pairing corresponds to a matching in $B$. Thus the probability of $p$ winning is $\frac{m_B}{e(2k, 2k+1)}$, where $m_B$ is the number of matchings in $B$ and $e(2k, 2k+1)$ is the number of different ways to pair $2k$ of the $2k+1$ candidates in the preround (which is straightforward to compute). Thus, evaluating $p$'s chances of winning in this election is at least as hard as counting the number of matchings in an arbitrary $B$, which is #P-hard. Moreover, because we can compute $p$'s chances of winning solely on the basis of properties 2a, 2b, and 2c, and by property 2d, these properties are maintained for any single additional vote, it follows that a manipulator cannot affect $p$'s chances of winning. Thus, CONSTRUCTIVE-MANIPULATION in this case simply comes down to computing $p$'s chances of winning, which is #P-hard as demonstrated. ∎

---

[4]#P is the class of problems where the task is to count the number of solutions to a problem in NP.

## 5.2 A broadly applicable reduction

In this subsection we present a single broadly applicable reduction which will satisfy the preconditions of Theorem 7 for many voting protocols, including all of the protocols discussed in this paper, thus proving them #P-hard to manipulate when the voting precedes the preround scheduling.

**Definition 3** *We label the following reduction $R_1$. Given a bipartite graph $B$ with the same number of vertices $k$ in both parts (labeled $1$ to $k$ in one part, $k+1$ to $2k$ in the other), we construct the following set of $12k^3 + 2k^2$ votes:*

- *$6k^3$ votes that rank the candidates $c_{k+1} \succ c_{k+2} \succ \ldots \succ c_{2k} \succ p \succ c_1 \succ c_2 \succ \ldots \succ c_k$;*
- *$3k^2$ votes that rank the candidates $p \succ c_k \succ c_{k-1} \succ \ldots \succ c_1 \succ c_{2k} \succ c_{2k-1} \succ \ldots \succ c_{k+1}$;*
- *$6k^3 - 3k^2$ votes that rank the candidates $c_k \succ c_{k-1} \succ \ldots \succ c_1 \succ c_{2k} \succ c_{2k-1} \succ \ldots \succ c_{k+1} \succ p$;*
- *For each edge $(i, j)$ in $B$ ($1 \leq i \leq k$, $k+1 \leq j \leq 2k$), one vote that ranks the candidates $c_i \succ c_j \succ p \succ c_1 \succ c_2 \succ \ldots \succ c_{i-1} \succ c_{i+1} \succ \ldots \succ c_k \succ c_{k+1} \succ c_{k+2} \succ \ldots \succ c_{j-1} \succ c_{j+1} \succ \ldots \succ c_{2k}$, and another one that ranks them $c_{2k} \succ c_{2k-1} \succ \ldots \succ c_{j+1} \succ c_{j-1} \succ \ldots \succ c_{k+1} \succ c_k \succ c_{k-1} \succ \ldots \succ c_{i+1} \succ c_{i-1} \succ \ldots \succ c_1 \succ p \succ c_i \succ c_j$ (i.e., the inverse of the former vote, apart from $c_i$ and $c_j$ which have maintained their order);*
- *For each pair $i, j$ without an edge between them in $B$ ($1 \leq i \leq k$, $k+1 \leq j \leq 2k$), one vote that ranks the candidates $c_j \succ c_i \succ p \succ c_1 \succ c_2 \succ \ldots \succ c_{i-1} \succ c_{i+1} \succ \ldots \succ c_k \succ c_{k+1} \succ c_{k+2} \succ \ldots \succ c_{j-1} \succ c_{j+1} \succ \ldots \succ c_{2k}$, and another one that ranks them $c_{2k} \succ c_{2k-1} \succ \ldots \succ c_{j+1} \succ c_{j-1} \succ \ldots \succ c_{k+1} \succ c_k \succ c_{k-1} \succ \ldots \succ c_{i+1} \succ c_{i-1} \succ \ldots \succ c_1 \succ p \succ c_j \succ c_i$ (i.e., the inverse of the former vote, apart from $c_j$ and $c_i$ which have maintained their order).*

We now have to show that this reduction satisfies the preconditions of Theorem 7. We start with the properties that are protocol-independent.

**Theorem 8** *$R_1$ satisfies properties 2b and 2c of Theorem 7 (under any protocol $P$, because these properties are independent of $P$), even with a single additional arbitrary vote.*

**Proof**: In the pairwise election between $p$ and any one of the $c_i$ with $k + 1 \leq i \leq 2k$, $p$ is ranked higher in only $4k^2$ votes, and thus loses the pairwise election. So property 2b is satisfied. For a pairwise election between some $c_i$ and $c_j$ ($1 \leq i \leq k$ and $k + 1 \leq j \leq 2k$), the first $12k^3$ votes' net contribution to the outcome in this pairwise election is 0. Additionally, the two votes associated with any pair $q, r$ ($1 \leq q \leq k$ and $k + 1 \leq r \leq 2k$) also have a net contribution of 0, if either $q \neq i$ or $r \neq j$. The only remaining votes are the two associated with the pair $i, j$, so $c_i$ wins the pairwise election by 2 votes if there is an edge $(i, j)$ in $B$, and $c_j$ wins the pairwise election by 2 votes otherwise. So property 2c is satisfied. Because both are satisfied with a "margin" of at least 2, a single additional vote will not change this. ∎

Finally, because property 2a is protocol-dependent, we need to prove it for our reduction on a per-protocol basis. This is what the following four theorems achieve.

### 5.3 Examples

**Theorem 9** $R_1$ *satisfies property 2a of Theorem 7 under the Plurality protocol. This holds even when there is a single additional arbitrary vote.*

**Proof**: If at least one of the $c_i$ with $k + 1 \leq i \leq 2k$ is not removed, $p$ can get at most $5k^2$ votes, whereas the lowest-indexed remaining candidate among the $c_i$ with $k + 1 \leq i \leq 2k$ will get at least $6k^3$ votes, so $p$ does not win. On the other hand, if all the $c_i$ with $k + 1 \leq i \leq 2k$ are removed, $p$ will get at least $6k^3 + 3k^2$ votes, which is more than half the votes, so $p$ wins. In both cases there is a "margin" of at least 2, so a single additional vote will not change this. ∎

**Theorem 10** $R_1$ *satisfies property 2a of Theorem 7 under the Borda protocol. This holds even when there is a single additional arbitrary vote.*

**Proof**: If at least one of the $c_i$ with $k + 1 \leq i \leq 2k$ is not removed, consider the highest-indexed remaining candidate among the $c_i$ with $k + 1 \leq i \leq 2k$; call it $h$. The first $12k^3$ votes will put $h$ at least $9k^3 - 3k^2$ points ahead of $p$. ($12k^3 - 3k^2$ of them rank $h$ above $p$, and the $3k^2$ others can give $p$ an advantage of at most $k$ each.) The $2k^2$ remaining votes can contribute an advantage to $p$ of at most $k$ each, and it follows that $h$ will still have at least $7k^3 - 3k^2$ more points than $p$. So $p$ does not win. On the other hand, if all the $c_i$ with $k + 1 \leq i \leq 2k$ are removed, then there are two groups of $6k^3 - 3k^2$ among the first $12k^3$ votes which (over the remaining candidates) are each other's exact inverses and hence have no net effect on the scores. Also, the last $2k^2$ votes, which are organized in pairs, have no net effect on the score because (over the remaining candidates) the votes in each pair are each other's exact inverse. The remaining votes all rank $p$ highest among the remaining candidates, so $p$ wins. In both cases the "margin" is big enough that a single additional vote will not change this. ∎

**Theorem 11** $R_1$ *satisfies property 2a of Theorem 7 under the Maximin protocol. This holds even when there is a single additional arbitrary vote.*

**Proof**: If at least one of the $c_i$ with $k + 1 \leq i \leq 2k$ is not removed, then in any pairwise election between such a candidate and $p$, $p$ will get at most $5k^2$ votes. However, the lowest-indexed remaining candidate among the $c_i$ with $k + 1 \leq i \leq 2k$ will get at least $6k^3$ votes in every one of its pairwise elections. So $p$ does not win. On the other hand, if all the $c_i$ with $k + 1 \leq i \leq 2k$ are removed, $p$ will get at least $6k^3 + 3k^2$ votes in every one of its pairwise elections, which is more than half the votes; so $p$ wins. In both cases there is a "margin" of at least 2, so a single additional vote will not change this. ∎

**Theorem 12** $R_1$ *satisfies property 2a of Theorem 7 under the STV protocol. This holds even when there is a single additional arbitrary vote.*

**Proof**: If at least one of the $c_i$ with $k + 1 \leq i \leq 2k$ is not removed, consider the lowest-indexed remaining candidate among the $c_i$ with $k + 1 \leq i \leq 2k$; call it $l$. $l$ will hold at least $6k^3$ votes as long as it is not eliminated, and $p$ can hold at most $5k^2$ votes as long as $l$ is not eliminated. It follows that $p$ will be eliminated before $l$, so $p$ does not win. On the other hand, if all the the $c_i$ with $k + 1 \leq i \leq 2k$ are removed, $p$ will hold at least $6k^3 + 3k^2$ votes throughout, which is more than half the votes; so $p$ cannot be eliminated and wins. In both cases there is a "margin" of at least 2, so a single additional vote will not change this. ∎

**Theorem 13** *In any of $RPRE + Plurality$, $RPRE + Borda$, $RPRE + Maximin$, and $RPRE + STV$, CONSTRUCTIVE-MANIPULATION is #P-hard.*

**Proof**: Immediate from the previous theorems. ∎

## 6 PSPACE-hardness when scheduling and voting are interleaved

In this section, we increase the complexity of manipulation one more notch, to PSPACE-hardness,[5] by interleaving the scheduling and vote elicitation processes.

We first discuss the precise method of interleaving required for our result. The method is detailed and quite complicated. Nevertheless, this does *not* mean that the interleaving should always take place in this particular way in order to have the desired hardness. If the interleaving method used for a particular election is (say, randomly) chosen from a wider (and possibly more naturally expressed) class of interleaving methods containing this one, our hardness result still goes through, as hardness carries over from the specific to the general. Thus, our goal is to find the most specific method of interleaving for which the hardness still occurs, because this gives us the most information about more general methods. We only define the method for the case where the number of candidates is a multiple of 4 because this is the case that we will reduce to (so it does not matter how we generalize the protocol to cases where the number of candidates is not a multiple of 4).

**Definition 4** $IPRE$ *proceeds as follows:*

---
[5]PSPACE is the class of problems solvable in polynomial *space*.

1. Label the matchups (a matchup is a space in the pre-round in which two candidates can face each other; at this point they do not yet have candidates assigned to them) 1 through $\frac{|C|}{2}$;
2. For each matchup $i$, assign one of the candidates to play in it, and denote this candidate by $c(i, 1)$. Thus, one of the candidates in each matchup is known.
3. For some $k$ which is a multiple of 4, for each $i$ with $1 \leq i \leq k$, assign the second candidate to play in matchup $i$, and denote this candidate $c(i, 2)$. Thus, we have $k$ fully scheduled matchups.
4. For each pair of matchups $(2i-1, 2i)$ with $i > \frac{k}{2}$, assign two more candidates to face the candidates already in these two matchups, and denote them $c((2i-1, 2i), 1)$ and $c((2i-1, 2i), 2)$. (Thus, at this point, all that still needs to be scheduled is, for each $i$, which of these two faces $c(2i-1, 1)$ and which $c(2i, 1)$.)
5. For $i = \frac{k}{2} + 1$ to $\frac{|C|}{4}$:
   • Randomly decide which of $c((2i-1, 2i), 1)$ and $c((2i-1, 2i), 2)$ faces $c(2i-1, 1)$, and which faces $c(2i, 1)$. Denote the former $c(2i-1, 2)$, the latter $c(2i, 2)$,
   • Ask all the voters whether they prefer $c(i - \frac{k}{2}, 1)$ or $c(i - \frac{k}{2}, 2)$. (We observe that, even if the number of already scheduled matchups is $k = 0$, the elicitation process trails behind the scheduling process by a factor 2.)
6. Elicit the remainder of all the votes.

One important property of this elicitation process is that the voters are treated symmetrically: when a query is made, it is made to all of the voters in parallel. Thus, no voter gets an unfair advantage with regard to knowledge about the schedule. Another important property is that the elicitation and scheduling process at no point depends on how the voters have answered earlier queries. Thus, voters cannot make inferences about what other voters replied to previous queries on the basis of the current query or the current knowledge about the schedule. These two properties guarantee that many issues of strategic voting that may occur with vote elicitation [Conitzer and Sandholm, 2002b] in fact do not occur here.

We are now ready to present our result.

**Theorem 14** *Given a voting protocol $P$, suppose that it is possible, for any Boolean formula $\phi$ in conjunctive normal form (i.e., a SAT instance) over variables $V = X \cup Y$ with $|X| = |Y|$ (and corresponding literals $L$), to construct in polynomial time a set of votes over a candidate set containing at least $\{p\} \cup C_L \cup \{c_y^1 : y \in Y\}$ with the following properties:*

- *(Property 3a) If we remove, for each $v \in V$, one of $c_{+v}$ and $c_{-v}$, $p$ would win an election under protocol $P$ against the remaining candidates if and only if for every clause $k \in K$ (where $K$ is the set of clauses in $\phi$), there is some $l \in L$ such that $c_l$ has not been removed, and $l$ occurs in $k$. This should hold even if a single arbitrary vote is added.*
- *(Property 3b) For any $x \in X$, $c_x$ and $c_{-x}$ are tied in their pairwise election after these votes.*
- *(Property 3c) For any $y \in Y$, $c_y$ and $c_{-y}$ are both losing their pairwise elections against $c_y^1$ by at least 2 votes (so that they will lose them regardless of a single additional vote).*

*Then CONSTRUCTIVE-MANIPULATION in $IPRE + P$ is PSPACE-hard (and PSPACE-complete if $P$ can be executed in polynomial space).*

**Proof**: Consider the following election under $IPRE + P$. Let the candidate set be the set of all candidates occurring in the votes constructed from $\phi$ (the "original candidates"), plus one dummy candidate for each of the original candidates besides the $c_{+v}$ and $c_{-v}$. To each of the constructed votes, add all the dummy candidates at the bottom; let the resulting set of votes be the set of the nonmanipulators' votes, according to which they will answer the queries posed to them. The manipulator has yet to decide on its strategy for answering queries. After step 4 (according to Definition 4) of $IPRE + P$ (up to which point the manipulator will not have had to make any decisions), let the situation be as follows:

- The number of already fully scheduled matchups is $k = \frac{|C|}{2} - 2|Y|$. In matchup $i$ ($1 \leq i \leq |X|$), $c_{+x_i}$ faces $c_{-x_i}$. In the remaining fully scheduled matchups, candidates not corresponding to a literal face a dummy candidate.
- Matchups $k + 2i - 1$ and $k + 2i$ ($1 \leq i \leq |Y|$) already have candidates $c_{+y_i}$ and $c_{-y_i}$ in them, respectively. The other two candidates to be assigned to these rounds are $c_{y_i}^1$ and a dummy candidate.

Thus, what will happen from this point on is the following. For $i$ ranging from 1 to $|X|$, first the protocol will schedule which of $c_{+y_i}$ and $c_{-y_i}$ face which of $c_{y_i}^1$ and the dummy candidate. The $c_l$ facing the dummy will move on, and the other will be defeated by $c_{y_i}^1$, by property 3c. Second, everyone will be asked which of $c_{+x_i}$ and $c_{-x_i}$ is preferred, and because the nonmanipulators will leave this pairwise election tied by property 3b, the manipulator's vote will be decisive. Thus, we can think of this as nature and the manipulator alternatingly giving the variables in $Y$ and $X$ respectively truth-values: $v$ is set to *true* if $c_{+v}$ survives, and to *false* if $c_{-v}$ survives. By property 3a it then follows that $p$ wins if and only if the resulting assignment satisfies all the clauses, i.e. is a solution to the SAT instance. Thus, the manipulator's strategy for setting variables should aim to maximize the chance of the SAT instance being satisfied eventually. But this is exactly the problem STOCHASTIC-SAT, which is PSPACE-complete [Papadimitriou, 1985].

If $P$ can be executed in polynomial space, the manipulator can enumerate all possible outcomes for all possible strategies in polynomial space, so the problem is also in PSPACE. ■

Because the preconditions of Theorem 14 are similar to those of Theorem 1, we can build on our previous reductions to apply this theorem to the well-known protocols.

**Theorem 15** *For each of $Plurality$, $Borda$, $Maximin$, and $STV$, there exists a reduction that satisfies properties 3a, 3b and 3c of Theorem 14. Thus, In any of $IPRE+Plurality$, $IPRE + Borda$, $IPRE + Maximin$, and $IPRE + STV$, CONSTRUCTIVE-MANIPULATION is PSPACE-complete.*

**Sketch of Proof**: We can modify the reductions from Section 4 to satisfy the preconditions of Theorem 14. This is done by adding in the $c_y^1$ in such a way as to achieve property 3c (ranking them just above their corresponding $c_y$ and $c_{-y}$ in slightly more than half the votes), while preserving property 3a (by ranking them as low as possible elsewhere). ∎

# 7 Conclusions

Voting is a general method for preference aggregation in multiagent systems, but seminal results have shown that all (nondictatorial) voting protocols are manipulable. One could try to avoid manipulation by using voting protocols where determining a beneficial manipulation is hard computationally. A number of recent papers study the complexity of manipulating existing protocols.

This paper is the first work to take the next step of *designing new protocols that are especially difficult to manipulate*. Rather than designing these new protocols from scratch, we instead showed how to *tweak* existing protocols to make manipulation hard, while leaving much of the original nature of the protocol intact. The tweak studied in this paper consists of adding one preround to the election, where candidates face each other one against one. The surviving candidates continue to the original protocol. Surprisingly, this extremely simple and universal tweak makes typical protocols hard to manipulate! The protocols become NP-hard, #P-hard, or PSPACE-hard to manipulate, depending on whether the schedule of the preround is determined before the votes are collected, after the votes are collected, or the scheduling and the vote collecting are interleaved, respectively. We proved general sufficient conditions on the protocols for this tweak to introduce the hardness, and showed that the most common voting protocols satisfy those conditions. These are the first results in voting settings where manipulation is in a higher complexity class than NP (presuming PSPACE $\neq$ NP).

# References

[Bartholdi and Orlin, 1991] John J. Bartholdi, III and James B. Orlin. Single transferable vote resists strategic voting. *Social Choice and Welfare*, 8(4):341–354, 1991.

[Bartholdi *et al.*, 1989] John J. Bartholdi, III, Craig A. Tovey, and Michael A. Trick. The computational difficulty of manipulating an election. *Social Choice and Welfare*, 6(3):227–241, 1989.

[Conitzer and Sandholm, 2002a] Vincent Conitzer and Tuomas Sandholm. Complexity of manipulating elections with few candidates. AAAI, pages 314–319, Edmonton, Canada, 2002.

[Conitzer and Sandholm, 2002b] Vincent Conitzer and Tuomas Sandholm. Vote elicitation: Complexity and strategy-proofness. AAAI, pages 392–397, Edmonton, Canada, 2002.

[Ephrati and Rosenschein, 1991] Eithan Ephrati and Jeffrey S Rosenschein. The Clarke tax as a consensus mechanism among automated agents. AAAI, pages 173–178, Anaheim, CA, 1991.

[Ephrati and Rosenschein, 1993] Eithan Ephrati and Jeffrey S Rosenschein. Multi-agent planning as a dynamic search for social consensus. IJCAI, pages 423–429, Chambery, France, 1993.

[Gibbard, 1973] A Gibbard. Manipulation of voting schemes. *Econometrica*, 41:587–602, 1973.

[Papadimitriou, 1985] C Papadimitriou. Games against nature. *Journal of Computer and System Sciences*, 31:288–301, 1985.

[Pennock *et al.*, 2000] David M Pennock, Eric Horvitz, and C. Lee Giles. Social choice theory and recommender systems: Analysis of the axiomatic foundations of collaborative filtering. AAAI, pages 729–734, Austin, TX, 2000.

[Satterthwaite, 1975] M A Satterthwaite. Strategy-proofness and Arrow's conditions: existence and correspondence theorems for voting procedures and social welfare functions. *Journal of Economic Theory*, 10:187–217, 1975.

[Valiant, 1979] Leslie Valiant. The complexity of computing the permanent. *Theoretical Computer Science*, 8:189–201, 1979.